\documentstyle[aps,preprint]{revtex}
\newcommand{\be}{\begin{eqnarray}}
\newcommand{\e}{\end{eqnarray}}
\newcommand{\ep}{\epsilon}
\newcommand{\pp}{\partial}
\tighten
\begin{document}
%\draft

\title{Quark Transversity Distribution in Perturbative QCD: Light-Front 
Hamiltonian Approach}
\author{{\bf Asmita Mukherjee{\thanks{email: asmita@theory.saha.ernet.in}}
 and Dipankar Chakrabarti{\thanks{email: dipankar@theory.saha.ernet.in}}}\\
Saha Institute of Nuclear Physics\\1/AF Bidhannagar, Calcutta 700064,
India}
\date{12 th February, 2001}
\maketitle
\begin{abstract}
To resolve the current ambiguity in the splitting function corresponding to
the quark transversity distribution $h_1(x)$, we calculate $h_1(x)$ for a
dressed quark in light-front Hamiltonian perturbation theory. Our result
agrees with the expected form of the splitting function found in 
the literature and disagrees with the recent calculation in \cite{kuhn}.
We emphasize the importance of quark mass in $h_1(x)$ in perturbative QCD
and show its connection with a part of $g_T$.
\end{abstract}
\vskip .2in
\centerline{PACS: 11.10.Ef, 12.38.-t, 13.60.Hb}
\vskip .2in
%%%%%%%%%%%%%%%%%%%%%%%%%%%%%%%%%%
\noindent{\bf 1. Introduction} 
%%%%%%%%%%%%%%%%%%%%%%%%%%%%%%%%%%
\vskip .2in

The quark transversity distribution $h_1(x)$ has received some 
attention lately as a recent calculation \cite{kuhn} of its corresponding 
splitting function
does not agree with the expected result found in the literature \cite{ar}.
The transversity distribution function does not appear in polarized
inclusive deep inelastic scattering in the limit of massless quark, because 
it corresponds to a flip of
quark chirality which is absent in DIS when quark mass is zero. $h_1(x)$ 
can be measured in
polarized Drell-Yan process. $h_1(x)$ was first discussed by Ralston and
Soper for this process \cite{ral}. Since then, there has been a lot of
theoretical investigations and also several attempts to experimentally 
measure the transversity distribution (see \cite{kuhn,iof} and references
therein). However, the effect of quark mass has not been discussed so far. 

The leading contribution of the transversity distribution is twist two and
it gives the difference of quark densities with eigenvalues ${1\over 2}$ and
$-{1\over 2}$ in the eigenstates of the
transverse Pauli-Lubanski spin operator. In order to calculate its
corresponding splitting function, Artru and Mekhfi analyzed the helicity
amplitude of hadron-hadron scattering in the t channel and used
Altarelli-Parisi evolution equation.  
The splitting function calculated by
them is given by in the limit of zero quark mass,
\begin{eqnarray}
P_{s_{\perp}}(x)={C_f\over 2}\left[{4\over(1-x)_+}-4+3\delta(1-x)\right],
\label{art}
\end{eqnarray}
where the color factor $C_f={N^2-1\over {2N}} $ for $SU(N)$.
The delta function comes from a counterterm which is the same as in the
helicity non-flip splitting function, $P_{qq}(z)$.

However in a recent work, Hermann, Kuhn and Schutzhold have calculated the
splitting function for $h_1$ and their result does not agree with Artru {\it et
al} \cite{kuhn}. They have started with the  forward scattering amplitude
$T_\mu$ which contribute to the nonsinglet transversity distribution. By
Wick expansion of the products of currents in the amplitude, they found
that $32$ Feynman diagrams contribute to the splitting function, although
considering various constraints on $T$, they showed that calculation of $8$
Feynman diagrams is sufficient for their purpose. 
They have taken the quark
mass to be zero. The singularity structure
of the splitting function is found by using a dispersion relation and the
splitting functions are obtained by doing inverse Mellin transformation. The
splitting function is given by,
\be
P_\perp(x)={C_f\over {(4\pi)^2}} \Big (4-{4\over {(1-x)_+}} \Big )
\e
and the corresponding anomalous dimension
\be
\gamma_n={C_f\over {(4\pi)^2}} \Big ({4\over {1+n}}+4S_n\Big ), 
\e
where $$S_n=\sum_{j=1}^n{1\over j}.$$
The splitting function differs from the earlier result by a delta function. 
They have also calculated $g_1(x)$ in the same method. $g_1(x)$ calculated
in this method agrees with the expected result in the literature. On 
comparing the two
calculations, they have shown that the splitting function for $g_1$ contains
a delta function that comes from inverse Mellin transformation. The
behavior at $x=1$ is usually fixed by assuming a sum rule, whereas no such sum rule
is there for $h_1(x)$. They pointed out that in the earlier works, the
behavior of $h_1$ at $x=1$ has been fixed by hand.

To resolve the above confusion generated by mutually contradictory results,
we calculate the transversity distribution in an entirely different method
based on light-front Hamiltonian technique. To uncover the physical picture
of $h_1(x)$, it is most useful to go to light-front formalism where the 
polarized
and unpolarized parton distributions in hard scattering can be expressed as
the light-cone Fourier transform of the matrix element of bilocal operators
in light-front gauge $A^+=0$ \cite{jaji}. In this method we can directly
calculate the structure function itself and hence we do not need to do any
inverse Mellin transformation as other formalisms need. Also, the role
played by the quark mass can be seen clearly in this picture. 
We then compare $h_1(x)$ with the transverse polarized structure function
$g_T$.
\vskip .2in
% The leading contribution to the transversity
%distribution is given by \cite{jaji},
%\be
%h_1(x)=
%\e      
  
%\end{document} 
%%%%%%%%%%%%%%%%%%%%%%%%%%%%%%%%%%%%%%%%%%%%%%%%%%%
%\documentstyle[aps,preprint]{revtex}
%\begin{document}
\noindent{\bf 2. Transversity Distribution 
$h_1(x)$ in Light-Front Hamiltonian Perturbation Theory}
\vskip.2in
%###################################################
The leading contribution to the transversity distribution function $h_1(x)$
in $A^+=0$ gauge is
given by\cite{jaji}
\begin{eqnarray}
h_1(x) = {1\over{P^+}}\int{ d\lambda\over{2 \pi}}e^{i\lambda x}\langle
PS_{\perp}|\psi_+^\dagger(0)\gamma_{\perp}\gamma_5\psi_+{(y^-)}
|PS_{\perp}\rangle.
\label{hx}
\end{eqnarray}
Here $\lambda={1\over 2}P^+y^-$ and $\psi=\psi^++\psi^-$, where
$\psi^{\pm}=\Lambda^{\pm}\psi$, the projection operators
$\Lambda^{\pm}={1\over 4} \gamma^{\pm} \gamma^{\mp}$. The above expression
differs from Jaffe and Ji by a factor of $\sqrt 2$ in the denominator
because of our different convention.  $P^\mu$ and $S^\mu$ are the total
momentum and polarization of the target state. The operator in $h_1(x)$ does
not involve the constrained field $\psi^-$, so
is a 'good' operator. Because of this, unlike the transverse structure
function $g_T$, $h_1$ has a simple parton interpretation. Also, here we are
calculating $h_1(x)$ which is a gauge invariant object, in the light-front
gauge $A^+=0$. In this gauge, the link operator involved in the bilocal
expression is unity. For simplicity, we have not considered the flavor
dependence.

The representation of the light-front $\gamma$-matrix we use here is the
following
\begin{eqnarray}
\gamma^0=\left[\begin{array}{cc}0&-i\\i&0\end{array}\right],~~~~~\gamma^3=
\left[\begin{array}{cc}0&i\\i&0\end{array}\right], ~~~~~\gamma^i=
\left[\begin{array}{cc}-i\stackrel{\sim}{\sigma}^i&0\\0&
i\stackrel{\sim}{\sigma}^i\end{array}\right]
\end{eqnarray}
where
$\stackrel{\sim}{\sigma}^1=\sigma^2,~~\stackrel{\sim}{\sigma}^2=-\sigma^1$.
The expression Eq.(\ref{hx}) in general 
is nonperturbative and the target is a bound
state. However, this expression can be used in pertubative calculation of
the splitting function corresponding to $h_1(x)$ in a straightforward manner  
by  replacing the target state by a dressed quark state  in 
light-front Hamiltonian perturbation theory. 

The transversely polarized state can be expressed in terms of helicity
states as,
\begin{eqnarray}
|PS_{\perp}\rangle={1\over\sqrt2}(|P\uparrow\rangle\pm|P\downarrow\rangle).
\label{ps}
\end{eqnarray} 
Without any loss of generality, we take the state to be polarized along $x$
direction. 
The polarization vector $S^\mu$ is defined as,
\be
S^\mu={1\over 2}{\overline u}\gamma^\mu \gamma^5 u
\e
where $u$ is the Dirac spinor. The components of the above polarization
vector can be calculated for both longitudinal and transverse polarization.
For longitudinal polarization, we get, $S^+=P^+$, $S^\perp=0$. However for
transverse polarization along $x$ direction, 
the components  are, $S^2=0$, $S^1=M$,
$S^-=2{P^1\over P^+}M$, $S^+=0$ where $M$ is the mass of the state. {\it It is clear
that one needs a massive state in order to have transverse polarization.}
 
A dressed quark state of total momentum $P$ and helicity $\sigma$
can be expanded in Fock space as,
\begin{eqnarray}
|P\sigma\rangle=\sqrt{\it N} &&\Big\{ b^{\dagger}_{\sigma}(P)|0\rangle
+\sum_{\sigma_2,\lambda_2}\int{dk_1^+d^2k_1^{\perp}\over
\sqrt{2(2\pi)^3k_1^+}}
\int {dk_2^+d^2k_2^{\perp}\over\sqrt{2(2\pi)^3k_2^+}}
\psi_2(P,\sigma|k_1,\sigma_2;k_2,\lambda_2) \nonumber \\&&  
\times\sqrt{2(2\pi)^3 P^+}\delta^3(P-k_1-k_2)b^{\dagger}_{\sigma_2}(k_1)
a^{\dagger}_{\lambda_2}(k_2)|0\rangle + ...\Big\}
\end{eqnarray}
where ${\it N}$ is the normalization constant 
and $\psi_2(P,\sigma|k_1,\sigma_2;k_2,\lambda_2)$ is the probability
amplitude to find a bare quark with momentum $k_1$ and helicity $\sigma_2$
and a bare gluon with momentum $k_2$ and helicity $\lambda_2$ in the dressed
quark. Here, we consider upto two-particle sector in the Fock-space.

We introduce the Jacobi momenta 
$$ x_i={k_i^+\over P^+}, ~~\kappa_i^{\perp}=k_i^{\perp}-x_iP^{\perp}$$ so
that $${\sum_i x_i} =1, ~~{\sum_i\kappa_i^{\perp}}=0.$$ 
We redefine $\psi_2$ in a boost invariant form
$$\sqrt{P^+}\psi_2(k_i^+,k_i^{\perp})=\Phi_2(x_i,\kappa_i^{\perp}).$$
The two particle amplitude is given by \cite{rajen}
\begin{eqnarray}
\Phi_2^{\sigma_2,\lambda_2}(x,\kappa^{\perp};1-x,-\kappa^{\perp})
=&&{1\over \left[ m^2-{m^2+{\kappa^{\perp}}^2\over x}-{{\kappa^{\perp}}^2
\over 1-x}
\right]}{1\over \sqrt{1-x}}T^a\chi_{\sigma_2}^{\dagger} \nonumber \\&&
\times\left[-2{\kappa^{\perp}\over{1-x}}-{\sigma^{\perp}\cdot\kappa^{\perp}
-im \over x}\sigma^{\perp}-\sigma^{\perp}im\right]\chi_{\sigma}\cdot
(\epsilon_{\lambda_2}^{\perp})^\star. 
\label{phi2}
\end{eqnarray} 
Here $m$ is the mass of the quark and $\chi_{\sigma}$ is the two component 
spinor for the quark and $\epsilon_{\lambda_2}^{\perp}$ is the gluon
polarization vector. 
%(ref: A. harindranath, R. kundu, W-M Zhang, PRD59, 094013, (1999))\\

We first calculate the matrix element contribution to $h_1(x)$ between two 
helicity states (of helicity $\sigma$ and $\sigma^{\prime}$) of dressed
quarks, 
\begin{eqnarray}
h^{\prime}(x)={1\over{P^+}}\int{ d\lambda\over{2 \pi}}e^{i\lambda x}\langle
P\sigma|\psi_+^\dagger(0)\gamma_{\perp} \gamma_5\psi_+(y^-)
|P\sigma^{\prime}\rangle.
\end{eqnarray}

Since the state is polarized along $x$ direction,
$\gamma_{\perp}=\gamma_1$. In the two component formalism 
\begin{eqnarray}
\psi_+(x)=\left[\begin{array}{c}\xi(x)\\0\end{array}\right]
\end{eqnarray}
and the chirality flip operator reduces to
\begin{eqnarray}
\psi_+^\dagger(0)\gamma_{\perp}\gamma_5\psi_+(y) =
-\xi^\dagger(0)\sigma^1\xi(y)
\end{eqnarray}
with
\begin{eqnarray}
\xi(x)=\sum_{\lambda}\chi_{\lambda}\int{dk^+d^2k^{\perp}\over 
2(2\pi)^3\sqrt{k^+}}[b_{\lambda}(k)e^{-ikx}
+d^{\dagger}_{-\lambda}(k)e^{ikx}].
\end{eqnarray}
At this stage,
 the mass terms in
the vertex Eq.(\ref{phi2}) contains $\sigma^\perp$ which flips helicity. These  terms give
zero contribution in $h_1$ because of the $\sigma^1$ present in the
operator.
We now have
\be
h^{\prime}(x)&&
={1\over{P^+}}\int{ d\lambda\over{2 \pi}}e^{i\lambda x}\langle
P\sigma|-\xi^\dagger(0)\sigma^1\xi(y)|P\sigma^{\prime}\rangle \nonumber \\
&&=-{\it N}\left[\delta(1-x)+{\alpha_sC_f\over 4\pi^2}{4x\over 1-x}\int
d^2\kappa^{\perp}{1\over(\kappa^{\perp})^2}\right](\chi_{\sigma}^{\dagger}
\sigma^1\chi_{\sigma^{\prime}}). 
\e 
The first term comes from the single particle sector and the second term
comes from the two-particle (one quark and one gluon) sector.
 Regulating the lower limit by $\mu$ and upper limit by Q in $\kappa$
integration
we get
\begin{eqnarray}  
h^{\prime}(x)=-{\it N}\left[\delta(1-x)+{\alpha_sC_f\over 2\pi}{2x\over
1-x}ln{Q^2\over \mu^2}\right](\chi_{\sigma}^{\dagger}
\sigma^1\chi_{\sigma^{\prime}}).
\end{eqnarray}
$h^{\prime}(x)\neq 0$ only if $\sigma \neq \sigma^{\prime}$.

The normalization of the state 
\be
\langle P\sigma|P^{\prime}\sigma^{\prime}\rangle = 2(2\pi)^3P^+\delta(P^+-
{P^{\prime}}^+)\delta^2(P^{\perp}-{P^{\prime}}^{\perp})\delta_{\sigma,
\sigma^{\prime}}
\e
gives,
\be
{\it N}\Big [1+{\alpha_s\over {2\pi}}C_f \int_\ep^{1-\ep} 
dx {1+x^2\over 1-x}ln{Q^2\over \mu^2} \Big ]=1.
\e
Upto order $\alpha_s$ we get \cite{rajen}
\begin{eqnarray}
{\it N}=1-{\alpha_s \over 2\pi}C_f\int_\ep^{1-\ep} dx {1+x^2\over 1-x}
ln{Q^2\over \mu^2}.
\end{eqnarray}
It is important to mention here that, the mass terms in the vertex in 
Eq.(\ref{phi2}) gives $m^2$ terms in the normalization condition, which are
power suppressed. We have neglected such contributions. In the denominator,
$m^2$ is neglected compared to $(\kappa^\perp)^2$ since
$m^2<<(\kappa^\perp)^2$.
Here $\ep$ is a small cutoff on the longitudinal momentum fraction $x$ which
can be safely taken to be zero at the end of the calculation.
Using the
definition Eq.(\ref{ps}) for the transversely polarized state in terms of the
helicity states, we get the
transversity distribution function
\begin{eqnarray}
h_1(x)=\delta(1-x)+{\alpha_s\over 2\pi}C_f ln {Q^2\over {\mu^2}} \Big [
{2x\over 1-x}-\delta(1-x)\int_\ep^{1-\ep} dy {1+y^2\over 1-y}\Big ]
\label{h1}
\e
which can be written as,
\be
h_1(x)=\delta(1-x)+{\alpha_sC_f\over 2\pi}\Big({2\over
(1-x)_+}-2+{3\over 2}\delta(1-x)\Big)ln{Q^2\over \mu^2}.
\end{eqnarray} 
 The splitting function can be easily obtained from the above expression: 
\begin{eqnarray}
P_{s_{\perp}}(x)={C_f\over 2}\left[{4\over(1-x)_+}-4+3\delta(1-x)\right]
\end{eqnarray} 
which exactly agrees with Eq.(\ref{art}). Depending on the plus or
minus sign in the definition of $|PS_\perp\rangle$ given by Eq.(\ref{ps}), 
the result can have   an overall minus sign. 
The anomalous dimension is defined as,
\be
\gamma_n= \int_0^1dz z^{n-1} P_{s_\perp}(z).
\e
Using the above splitting function, we get,
\be
\gamma_n= C_f \Big ({3\over 2}-2\sum_{j=1}^n {1\over j} \Big )
\e
which again agrees with the expected result found in the literature
\cite{ar,bar}. However, the above analysis is done is a completely different
approach based on light-front Hamiltonian perturbation theory. Instead of
Feynman diagrams, here we have old-fashioned $x^+$ ordered diagrams. The
complexities of the operator product expansion are absent. There is no need
to calculate the moments of the structure functions and then invert it. In
this formalism, we get the expressions of the structure functions themselves
directly in terms of bilocal matrix elements. As we have shown above, the 
splitting functions can be obtained easily by calculating the matrix element 
for a dressed parton state. In our calculation, 
we have not assumed any sum
rule for $h_1(x)$. The final result for the splitting function is obtained
only after taking the normalization of the state into account.

Another very important point is that, the role played by the quark mass in
the perturbative analysis of $h_1(x)$ can be seen very clearly in this picture, whereas it is obscure in
other formalisms. In order to show clearly the effect of quark mass, we
compare our calculation with the calculation of $g_T$ in this method.
  
\vskip .2in
%%%%%%%%%%%%%%%%%%%%%%%%%%%%%%%%%%%%%%%%%%%%%%%%%%%%%%%%%%%%%%%%%%
\noindent{\bf 3. Comparison with the Transverse Polarized Structure Function $g_T$}
%%%%%%%%%%%%%%%%%%%%%%%%%%%%%%%%%%%%%%%%%%%%%%%%%%%%%%%%%%%%%%%%%
\vskip .2in
It is very interesting to compare and contrast the above calculation with the
calculation of $g_T$ for a dressed quark in order to understand the effect
of quark mass in the two cases. Recently, $g_T$ has been
calculated in light-front Hamiltonian perturbation theory \cite{wan}. $g_T$
 is given by,
\be
g_T(x,Q^2)={1\over {8\pi(S_\perp-{P_\perp\over P^+}S^+)} } \int d\eta 
		e^{-i\eta x}  \langle PS| (O_m+O_{k_\perp}+O_g)+h.c. 
                |PS \rangle 
\e
where
\be
O_m&&=m\psi_+^\dagger (\xi^-)\gamma_\perp {1\over {i\pp^+}}\gamma_5
\psi_+(0),\\
O_{k_\perp}&&=-\psi_+^\dagger(\xi^-) (\gamma_\perp {1\over
{\pp^-}}\gamma^\perp \cdot \pp^\perp  +2{P_\perp\over P^+})\gamma_5
\psi_+(0),\\	    
O_g&&=g\psi_+^\dagger(\xi^-)(\gamma^\perp \cdot A^\perp(\xi^-)){1\over
{i\pp^+}}\gamma_\perp\gamma_5 \psi_+(0),
\e
and $h.c.$ stands for hermitian conjugate.
In the above expression, we have used the equation for the constrained field
$\psi^-$. The operator $O_m$ involves $\gamma^\perp \gamma^5$ which is
similar to $h_1$. In the dressed quark calculation, the mass terms in the
expression of $\Phi_2$ do not contribute in $O_m$, exactly like $h_1$.
However, the other parts of the operator involve $\gamma^\perp \gamma^\perp
\gamma^5$  which gets nonzero contribution only from the helicity flip (mass
dependent) part of the vertex. So in the case of $g_T$, the mass of the
quark plays a very important role both in the transversely polarized state
and in the matrix element but in $h_1$, though the mass terms in the vertex
do not contribute in the matrix element, quark mass is still important in
perturbative calculations because one
cannot get a transversely polarized dressed quark state in the massless theory. We
emphasize that this effect can be seen more clearly in our approach, because
it is a Hamiltonian perturbation theory approach and the state plays a very 
important role.

The contribution of the mass dependent part $O_m$ to $g_T$ for a dressed
quark is given by \cite{wan}:
\be
g^m_T={1\over 2}\Big [ \delta(1-x)+{\alpha_s\over {2\pi}} C_f ln{Q^2\over
\mu^2} \Big ( {2\over 1-x}-\delta(1-x)\int_\ep^{1-\ep} dy{1+y^2\over 1-y}\Big )
 \Big ].
\e
The above expression is similar to $h_1$ Eq.(\ref{h1}), the additional $x$
in the numerator of the second term  in Eq.(\ref{h1}) comes due to the
absence of the ${1\over \pp^+}$ in the operator. Thus in perturbation theory,
we directly see the connection of $h_1(x)$ with the mass dependent part of
$g_T$. 
\vskip .2in
%%%%%%%%%%%%%%%%%%%%%%%%%%%
\noindent{\bf 4. Conclusion}
%%%%%%%%%%%%%%%%%%%%%%%%%%%%%%%
\vskip .2in
We have calculated the splitting function corresponding to the transversity
distribution upto $O(\alpha_s)$ in a recently developed light-front Hamiltonian
QCD approach. The
structure function in hard scattering can be directly expressed in terms of
the matrix elements of light-front bilocal currents. There is no need to do
an inverse Mellin transformation and the splitting functions can be easily
extracted. Another advantage over the Altarelli-Parisi method is that here one
deals with the probability amplitudes instead of probability densities and
real and virtual processes are calculated to the same order without any
difficulty. Here we have not assumed any particular behavior of $h_1(x)$ at
$x=1$. The delta function with the correct coefficient in the final
expression comes from  the normalization constant of the state. 

We have
taken the quark mass to be nonzero in our calculation and shown that $h_1(x)$ can be
related to one part of the transverse polarized structure function $g_T$. 
We have emphasized the importance of quark mass in the perturbative analysis
of $h_1(x)$. To our
knowledge, this is done for the first time in the literature. We emphasize that
in light-front Hamiltonian perturbation theory, this effect is seen 
in a more clear way since it is not a Feynman diagram based calculation.

Our result not only verifies the earlier result of Artru et. al. \cite{ar}
but does so in a completely different and much more straightforward way. 
The disagreement with \cite{kuhn}
introduces questions regarding the validity of various assumptions made in
their paper. 
\vskip .2in
      We acknowledge Prof. A. Harindranath for suggesting this problem and
also for many useful discussions.
%%%%%%%%%%%%%%%%%%%%%%%%%%%%%%%%%%%%%%%%%%%%%%%%%%%%%%%%%%%%%%%%%%%%%%%%%%

\end{document}